# Mode substitution induced by electric mobility hubs: results from Amsterdam

**Fanchao Liao**\*
Radboud University
Nijmegen School of Management
Nijmegen, the Netherlands
E-mail: fanchao.liao@ru.nl

**Jaap Vleugel**
Delft University of Technology
Faculty of Civil Engineering and Geosciences
Department of Transport & Planning
P.O. Box 5048, 2600 GA Delft, the Netherlands

**Gustav Bösehans**
Newcastle University
School of Engineering
NE1 7RU Newcastle upon Tyne, United Kingdom

**Dilum Dissanayake**
University of Birmingham
School of Geography, Earth and Environmental Sciences
B15 2TT Birmingham, United Kingdom

**Neil Thorpe**
Newcastle University
School of Engineering
NE1 7RU Newcastle upon Tyne, United Kingdom

**Margaret Bell**
Newcastle University
School of Engineering
NE1 7RU Newcastle upon Tyne, United Kingdom

**Bart van Arem**
Delft University of Technology
Faculty of Civil Engineering and Geosciences
Department of Transport & Planning
P.O. Box 5048, 2600 GA Delft, the Netherlands

**Gonçalo Homem de Almeida Correia**
Delft University of Technology
Faculty of Civil Engineering and Geosciences
Department of Transport & Planning
P.O. Box 5048, 2600 GA Delft, the Netherlands

*: Corresponding author*

**Abstract**

Electric mobility hubs (eHUBS) are locations where multiple shared electric modes including electric cars and e-bikes are available. To assess their potential to reduce private car use, it is important to investigate to what extent people would switch to eHUBS modes after their introduction. Moreover, people may adapt their behaviour differently depending on their current travel mode. This study is based on stated preference data collected in Amsterdam. We analysed the data using mixed logit models. We found users of different modes not only have a varied general preference for different shared modes, but also have different sensitivity for attributes such as travel time and cost. Compared to car users, public transport users are more likely to switch towards the eHUBS modes. People who bike and walk have strong inertia, but the percentage choosing eHUBS modes doubles when the trip distance is longer (5 or 10 km).

# 1. Introduction

Mobility hubs have recently gained popularity in many cities (Anderson et al., 2017; Chauhan et al., 2021; Kelley et al., 2020; Miramontes et al., 2017; Roukouni et al., 2023). They are defined as a location where multiple shared transport modes (e.g., vehicles, bikes, cargo-bikes, scooters) are provided. Making these mobility services shared and available at hubs is expected to deliver substantial benefits for the citizens and cities. On the one hand, it can potentially mitigate the negative externalities such as $CO_2$ emissions, air pollution, road congestion and high demand for parking space brought about by road transport which is mostly done by private cars (Becker et al., 2018; Kondor et al., 2019; Le Vine and Polak, 2019; Tikoudis et al., 2021). On the other hand, the provision of new transport alternatives also is supposed to enhance accessibility for locations that were not previously well connected thus promoting transport equity for people living in these regions (Guo et al., 2020; Jiao and Wang, 2021; Portland Bureau of Transportation, 2019; Shaheen et al., 2017). Electric mobility hubs or eHUBS which are equipped with EVs and e-bikes are expected to strengthen the potential for combatting pollution since electric modes have zero tailpipe related emissions (Millard-Ball et al., 2005; Shaheen and Cohen, 2013; Vasconcelos et al., 2017) although particles due to road abrasion and braking remain.

For the eHUBS' potential in reducing externalities to materialise, sufficient mode substitution needs to occur, in particular the replacement of private car trips, which can eventually lead to a reduction of private car ownership (Ye et al., 2021). In order to assess the potential impact of eHUBS, knowledge regarding people's adoption and usage of the shared electric modes in eHUBS is crucial. More specifically, it is important to know to what extent people would use eHUBS to replace the modes they are currently using. Whilst substituting car trips is the desired impact, eHUBS also may substitute the use of public transport and active modes. If this latter substitution occurs to a greater extent than compared to the former, it may even lead to a net increase in terms of emission which is unwanted. Furthermore, since it is already known that people's choice of current mode, and their subjective perceptions of it, significantly affects their preference for new mobility services (Mo et al., 2021), it will also enhance our understanding of eHUBS adoption by studying how eHUBS preference and substitution patterns vary for current users of different modes.

There has been much previous research regarding the intention and choice for adopting car-sharing. Many researchers have explored the general intention to use this mode, while several studies have conducted stated choice experiments to also examine the impact of trip and mode attributes on carsharing behaviour (Jin et al., 2020; Kim et al., 2017). Much less research focused on consumer preferences for shared e-bikes (Reck et al., 2022; Teixeira et al., 2021). Lately there have been some studies using GPS tracking data or operational data from shared mobility service providers to study factors impacting the demand for existing shared e-bike services (Reck et al., 2022, 2020); however, these systems are mostly dockless which are expected to be different from docked systems and some attributes such as travel cost (unit cost of shared services) cannot be varied and sometimes are unknown. Moreover, among these studies on shared mobility, only a limited number applied the results to specifically demonstrate the substitution pattern between modes (Li and Kamargianni, 2020; Papu Carrone et al., 2020), and did not account for the influence of the current modes being used by respondents.

In order to better understand the potential of eHUBS modes in substituting other transport modes and reducing negative externalities, we aim to answer the following research questions:

- How do mode and trip attributes affect people's decision on using eHUBS modes (shared electric cars and e-bikes) to substitute their current modes?
- How does the pattern of mode substitution vary depending on people's current modes?

In order to answer these questions, we conducted a survey including a stated choice experiment regarding people's choice of whether or not to use eHUBS modes to replace their current mode of transport. We have estimated multiple choice models to analyse the data; furthermore, we derived insights concerning mode substitution patterns and policy impact by applying the modelling result as input for scenario simulation.

The remainder of this paper is organised as follows: Section 2 gives a critical overview of the existing literature on shared electric mobility adoption and its mode substitution; Section 3 presents the methodology for data collection and analysis including survey design and model specification; Section 4 discusses the results of the modelling. The final section draws conclusions and discusses implications, limitations, and potential for future research.

## 2. Literature review

There are several different approaches to study how shared EVs and e-bikes substitute existing modes and especially their potential to reduce car use. The first approach investigates how the aggregate use of transport modes of each individual (instead of at the individual trip level) changes after adopting a new mode. Kopp et al. (2015) compared the demand for different transport modes between members of carsharing services and non-members. Other studies use longitudinal data to study whether e-bike usage substitutes car use (Haas et al., 2021; Kroesen, 2017; Sun et al., 2020). Their approach applied regression or cross-lagged panel models to examine whether the e-bike adoption/use reduces car use (and also affects the use of other modes). They found that e-bike adoption and use significantly reduced the number of conventional bike trips in the Netherlands, while it only substituted car and public transport trips to a lesser extent, which may lead to an unwanted aggregate impact. These findings can only provide limited insights into our research questions. Firstly, these studies were only concerning private e-bikes, probably because shared e-bikes are a fairly recent innovation and rendering the collection of longitudinal data impossible. As previous studies have found that shared EVs are used in a different way compared to privately owned ones (Wang et al., 2020), this difference between shared services and own vehicle also may be present for e-bikes (Reck et al., 2022; Teixeira et al., 2021). Secondly, the studies only used aggregate indicators such as the number of trips by a certain mode, which does not allow the investigation of mode substitution at the trip-level.

The second type of study directly asks current car/bike-sharing users which mode would they have used for their trips had the sharing service not been available. Liao and Correia (2020) reviewed several studies using this method to learn the mode substitution effect for shared EVs and e-bikes (Becker and Rudolf, 2018; Campbell et al., 2016; Martin and Shaheen, 2016). Bigazzi and Wong (2020) conducted a meta-analysis of 24 studies and found that the median mode substitution is the highest for public transit (33%), followed by conventional bicycle (27%), automobile (24%), and walking (10%). It included studies on shared and private e-bikes and shared e-bikes are found to substitute less car trips compared to private e-bikes. More recent studies in the US however found that the mode substitution for public transit is much lower (5%) (Fitch et al., 2020; Fukushige et al., 2021). The common limitation of this type of study is that it only provides descriptive statistics (X% of shared EV/e-bike trips would have been conducted using Y mode). Zhou et al. (2023) categorized trips by different distances and calculated a distinct substitution pattern for each distance. Fukushige et al. (2021) went a step further by estimating a statistical model: it uses variables including trip attributes and individual characteristics to predict the substituted mode of each shared e-bike trip. However, it did not include attributes of different modes such as travel time and cost.

The third approach analyses data of shared mobility trips. These trip data can be directly analysed on an aggregated level via spatial regression to find out the impact of factors such as land use and population density on shared mobility demand (Guidon et al., 2020, 2019). If data from multiple shared mobility services are available, a choice model between different shared mobility services can also be estimated (Reck et al., 2020). However, these studies do not provide information regarding how shared service substitute traditional modes such as private car and public transport. (Reck et al., 2022) collected a large dataset which enabled the estimation of a mode choice model between both traditional modes (private car, public transport) and shared mobility including an existing shared e-bike service. But this study can only provide limited insight for our research questions due to two reasons. First, the shared e-bike service has a dockless system which is expected to have different characteristics from docked systems such as eHUBS: the common trip distance range for dockless shared e-bike system is 1-3.5km (Guidon et al., 2019; Reck et al., 2020), while the average distance of docked systems is 4.2 km (Bielinski et al., 2021) or even around 7.5km (He et al., 2019) and a non-trivial percentage of trips are even longer than 10km (He et al., 2019). The longer distance of docked shared e-bike trips indicates that it may have larger potential to replace more car trips. Second, these services in running do not allow adjustment in level-of-service attributes such as unit price of shared mobility, which makes it impossible to explore the impact of these attributes on mode choice. The studies above used data from existing e-bike sharing systems and their users: apart from the limitations already discussed above, the results of these studies may also suffer from sample self-selection: since shared electric mobility services are not yet widely available, individuals who are already using these services are early adopters and may have different preferences compared to the other sectors of the population. In order to explore the potential of shared e-mobility services among the general population and investigate the impact of level-of-service attributes such as travel time and cost, many studies used the stated preference approach. Some of these studies do not provide insights on mode substitution due to limitations such as only eliciting the general intention of using the new mode without trip and service attributes (Efthymiou et al., 2013; Efthymiou and Antoniou, 2016), focusing on services with a specific purpose (park & carsharing, Cartenì et al., 2016) or studying membership choice instead of mode choice (Le Vine et al., 2014). As mentioned in the introduction, several studies investigated people's preference for carsharing and explicitly focused on its implication regarding mode substitution (Li and Kamargianni, 2020; Papu Carrone et al., 2020). Many factors were found to have significant impacts on mode choice and substitution, including mode attributes (such as travel time, travel cost, parking time and cost), trip contexts (trip purpose, trip distance) and individual characteristics (socio-demographic variables and attitudes). There were however few stated choice experiments involving shared e-bikes (or e-bikes in general). For example, Li and Kamargianni (2018) included private e-bike as an alternative in their choice experiment, although it does not form the focus of their analysis. Moreover, most of these studies did not examine how the substitution pattern differs between people currently using different modes.

As for the impact of current mode on people's mode choice and substitution patterns involving innovative modes such as electric and shared mobility, there have only been a few relevant studies. Bielinski et al. (2021) found that people who reduced their public transport use are more likely to use shared e-bike. Ceccato and Diana (2021) analysed substitution on the trip level. They asked people whether they were willing to use car-sharing for a certain trip (chain) in their travel diary and found that the intention is affected by the modes they previously used for this trip. Since people can state multiple previously used modes for a specific trip, the lack of information regarding the frequency of using each mode may bias the analysis. Furthermore, the propensity of using carsharing for a trip is not entirely equivalent to preference over the mode they are currently using. In a study of a different innovative mode (autonomous mobility-on-demand service, AMOD), Mo et al. (2021) incorporated the interaction of people's current modes with the Alternative Specific Constant (ASC) of AMOD and found all of them to be significant, which reveals the impact of existing modes on the preference for new modes. However, they did not investigate whether users of different current modes also have different sensitivity to the attributes of the new mode. De Luca and Di Pace (2015) explicitly addressed this when analysing stated preference data for a longer-distance (>10 km) inter-urban car-sharing programme. The authors have estimated conditional switching models separately for users of private cars, car-pooling and buses; users of these three modes indeed have different sensitivities for car-sharing service attributes. Based on indicators of model fit and elasticity values, this modelling approach also is found to be better than assuming a universal choice set for all individuals.

To summarize, despite the existing contribution regarding substitution between (shared) electric modes and existing modes, there are two main research gaps: first, there is no research investigating the mode choice and substitution pattern involving docked shared e-bike services; second, most studies did not explicitly account for the influence of the mode currently used by people on their choice regarding using shared electric modes.

Given the identified research gaps, this study contributes to the existing literature in the following ways. Firstly, we explore to what extent people will use shared EVs and e-bikes in eHUBS to replace their current modes for short trips. More specifically, we investigate how mode (such as travel time and cost) and trip (e.g. purpose) attributes affect the substitution by conducting trip-level analysis. Secondly, we specifically focus on how the preference for eHUBS modes and the substitution pattern vary for people who currently use different modes.

## 3. Methods of data collection and analysis
### 3.1 Survey design

We conducted an online survey to collect stated preference data and investigate people's mode choices when the shared electric mobility services become available. The survey consisted of four sections:

- respondents' current mobility profile including their ownership of vehicles and current travel pattern;

- information regarding typical trips for a given purpose;
- stated choice experiments for mode choice; and
- sociodemographic information.

The eHUBS service in our stated choice experiments is assumed to be a one-way station-based system: the users can pick up a vehicle from an eHUB station and return it to any other station in the same city. We choose this setting because it does not involve the consideration of activity duration at the destination. We also assume that a shared vehicle is always available in eHUBS and there is no parking search time when returning the shared vehicle. To ensure that all respondents have some basic knowledge of eHUBS, they were given an introduction to the basic characteristics and procedures of using an eHUB before the start of the experiment.

Since people's mode choice may be different depending on their trip purpose, we designed two separate experiments: one for commuting and the other for non-commuting trips. Respondents were directed to the commuting experiment if their current commute trip meets the following criteria: 1) no longer than 10 km (longer trips are likely inter-city and unlikely to be covered by the same sharing service network); 2) at least 3 times per week (we prefer to obtain responses from more frequent and experienced commuters) and 3) not by walking (we assume these trips are most likely rather short and very few would intend to switch towards eHUB). All respondents completed the non-commuting experiment.

For non-commuting experiments, we included multiple contexts aiming to cover different trip distances and purposes. Regarding distance, each choice task is concerning a trip of around 2, 5 or 10 km. Mode shares are expected to be different for these three distinct distance ranges: 5 km is no longer within "walking distance" for most people; trips over 10km are no longer considered "short trips" and are unlikely to be covered by an intra-city shared mobility service. A similar division was adopted in a previous mode choice study (Li and Kamargianni, 2020).

We also covered both leisure and shopping trips since shopping trips involve the transport of goods which enables us to explore the preference for electric cargo-bikes.

We adopted a design similar to a stated adaptation experiment (see Langbroek et al., 2017; Pan et al., 2019 for recent studies using this approach) instead of including all viable transport modes as alternatives in each choice task. In all choice tasks, the respondents chose between three alternatives: their current mode, shared EVs in an eHUB and shared e-bikes in an eHUB (or shared e-cargo bike in case of a shopping trip). The current mode alternative is thus respondent-specific. Not only does this reduce the cognitive burden of respondents, but it also presents a more realistic choice problem. Our focus is not to study the substitution between existing modes due to the change in their level of service. Given the *status quo*, we assume that a respondent's currently used mode is his/her most preferred among all existing modes; we aim to investigate to what extent people would switch from their currently most preferred mode to shared modes in eHUBS after its introduction. Figure shows an example choice task.

In the commuting experiment, we fix the attribute values of the current mode alternative as the real values provided by the respondent. The respondents provide the following details regarding their regular commuting trip (before COVID): approximate distance, most frequently used mode, travel time, congestion, travel cost, access and egress time. For those whose current mode is a private vehicle, we also ask for their parking search time and parking cost. The travel time of the eHUBS modes (excluding access and egress time) is calculated based on the distance of this reference trip. The design with a fixed reference alternative has been applied in many stated preference studies in transport (De Luca and Di Pace, 2015; Hensher et al., 2011; Krueger et al., 2016; Rose et al., 2008; Yan et al., 2019).

In the case of the non-commuting experiment, because we cover trips of two purposes (shopping and leisure) and three trip distances, it is difficult to elicit all detailed information for six reference trips. Therefore, we only ask for the respondent's default mode for a typical trip of a given distance and purpose (for example a 2 km leisure trip). Since each trip may not have the same destination unlike the case of commuting (e.g. a respondent may conduct several 5km leisure trips to different locations), the attributes of the current mode are also varied according to the experiment design as the other two eHUBS alternatives.

All attributes involved in the experiment design are varied by three levels. The attribute levels are based on their current value range and/or values in possible future scenarios; in both cases, they shall be easily understood by the respondents. Table 1 and Table 2 list the attribute levels in the commuting and non-commuting experiment respectively.

For both commuting and non-commuting experiments, we constructed an orthogonal design consisting of 27 choice tasks. We did not adopt an efficient design because we do not yet have good estimates for the coefficients of eHUBS attributes and D-efficient design is not robust when priors are far from the real value (Walker et al., 2018). For the commuting experiment, each respondent is randomly assigned to six choice tasks. In the case of the non-commuting experiment, one choice task is randomly selected for each combination of trip distance (2, 5 and 10km) and purpose (shopping and leisure). Therefore, each respondent also gets six (3x2) tasks in total.

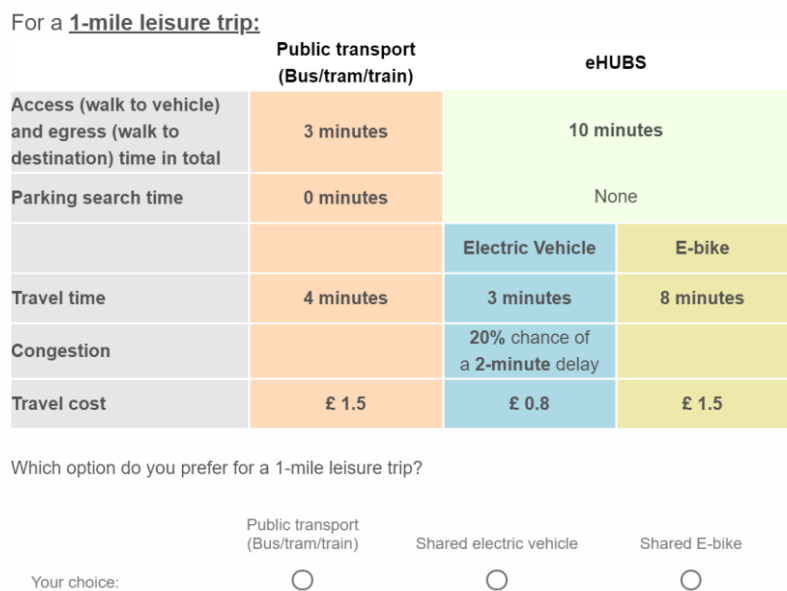

**Figure 1. Example of a choice task**

## Table 1 Attributes for the commuting experiment

| Attributes | eHUB | |
|---|---|---|
| **Access and egress time** | 2,10,18 mins | |
| | **EV** | **E-bike** |
| **Travel time** | If the current mode is private car: same as car<br>Otherwise:<br>Reference *80%,100%,120%<br>Reference is calculated based on distance assuming 30km/h | Reference * 80%,100%,120%<br>Reference is calculated based on distance assuming 20km/h |
| **Travel cost** | 0.15, 0.25, 0.35 €/min | €0.5, 1.5, 2.5 (regardless of distance)[1] |
| **Congestion level** | If current mode is private car: same as car<br>Otherwise:<br>Chance of delay: 0%, 20%, 40%<br>Possible delay: 25%, 50%, 75% of travel time | |

## Table 2 Attributes for the non-commuting experiment

| Attributes | eHUB | |
|---|---|---|
| **Access and egress time** | 2km: 2,6,10 min.<br>5km: 2,10,18min<br>10km: 2,10,18min | |
| | **EV** | **E-bike** |
| **Travel time** | 2km: 3, 5, 7 min<br>5km: 7, 10, 13min<br>10km: 15, 20, 25min | 2km: 4, 6, 8min.<br>5km: 10, 12, 14min<br>10km: 20, 25, 30min |
| **Travel cost** | 0.15, 0.25, 0.35 €/min | €0.5, 1.0, 1.5 (regardless of distance) |
| **Congestion level** | Chance of delay: 0%, 20%, 40%<br>Possible delay: 25%, 50%, 75% of travel time | |

| Attributes | Private car | PT | Private Bike | Walking |
|---|---|---|---|---|
| **Access and egress time** | Egress<br>2km: 1,3,5 min.<br>5km: 1,5,9 min<br>10km: 1,5,9 min | Egress<br>2km: 1,3,5 min.<br>5km: 1,5,9 min<br>10km: 1,5,9 min | | |
| **Travel time** | Same as shared EV | 2km: 4, 6, 8min<br>5km: 10, 15, 20min<br>10km: 20, 30, 40min | 2km: 6, 8, 10min<br>5km: 15, 20, 25 min<br>10km: 30, 40, 50min | 2km: 20, 25, 30min<br>5km: 50, 60, 70min |
| **Travel cost** | €0.1, 0.2, 0.3/km | 2km: 0.5, 1, 1.5 euro.<br>5km: 1, 1.5, 2 euro<br>10km: 2, 3,4 euro | | |
| **Congestion level** | Same as shared EV | | | |
| **Parking search time** | 0, 5, 10 min | | 0, 5, 10 min | |
| **Parking fee** | €0, 3, 6 | | | |

---

[1] The e-bike cost levels are not distance-specific because the current range of attributes can cover all distances explored in our study (less than 10km).

## 3.2 Data collection and sample characteristics

The target population for our survey was adults who hold a driver's licence and live in Amsterdam, in the Netherlands. We exclude people who do not have a driving license because our study involves shared electric vehicle service and these people are unable to use such a service. We collaborated with a panel company to distribute online surveys and collect responses in March 2021. We obtained 1003 responses. After excluding all responses with less than a 5-minute completion time (Because it is not reasonable to finish the survey within 5 minutes given the number of questions), the size of the valid sample used in the analysis is 880. All respondents completed the non-commuting experiment, while only 487 respondents participated in the commuting experiment because the others do not meet the criteria discussed in section 3.1.

**Table 3 Sample characteristics**

| Variable | Value | Percentage in sample | Percentage in population |
|---|---|---|---|
| **Gender** | Female | 54.1% | 50.4% |
|  | Male | 45.9% | 49.6% |
| **Age** | 18-24 | 16.0% | 12.0% |
|  | 25-34 | 27.4% | 26.1% |
|  | 35-44 | 19.3% | 17.4% |
|  | 45 or older | 37.3% | 44.5% |
| **Education** | No higher education | 39.5% | 52.0% |
|  | With higher education | 60.5% | 48.0% |
| **Household Income** | Low (<=€40.000) | 41.0% |  |
|  | Middle (>€40.000 and <=€80.000) | 32.7% |  |
|  | High (>€80.000) | 13.5% |  |
|  | Missing value | 12.8% |  |
| **Occupation** | Employed | 71.1% |  |
|  | Student | 8.4% |  |
|  | Others | 20.5% |  |
| **Number of children** | 0 | 63.9% |  |
|  | 1 | 16.4% |  |
|  | More than 1 | 19.7% |  |
| **Cars in household** | 0 | 17.5% |  |
|  | 1 | 65.0% |  |
|  | More than 1 | 17.5% |  |

Table 3 presents an overview of the socio-demographic variables of the sample which shows that the age distribution is broadly similar to the population, while males are slightly underrepresented. Highly-educated people are overrepresented probably because they are more accessible via online surveys. This shall be taken into account if one wishes to directly extrapolate the results to the entire population. We have decided not to adjust sample weight during the estimation of choice models as our main aim is not forecasting.

Table 4 illustrates the mode share of the *status quo* in both experiments. Although the majority of respondents have at least one car in their household, the share of private car use for short trips below 10km is consistently less than half. We notice that several modes (such as shared mobility) are used by only a small share of respondents (less than 10% or even 5%), while they have vastly different substitution patterns compared to other more commonly used modes: a much larger share of these people choose to switch to shared EV and e-bikes compared to users of other modes, but the absolute number of users of these modes is too small and does not allow us to estimate their taste parameters with sufficient reliability. Therefore, we choose to analyse only the substitution for those current modes that are used by a considerable number of respondents, namely private car, public transport, bike and walk in the non-commuting experiment. As for commuting trip experiment, only private car and bike are included. After this step, the number of remaining respondents in the non-commuting and commuting experiment are respectively 867 and 345.

**Table 4 The share of currently used modes for each trip context**

| Mode | Share in commuting trip | Share in 2km non-commuting trip | Share in 5km non-commuting trip | Share in 10km non-commuting trip |
|---|---|---|---|---|
| Private car | 38.4% | 16.2% | 28.6% | 48.0% |
| Car passenger | 2.1% | 4.6% | 5.3% | 7.0% |
| Public transport | 15.6% | 5.7% | 11.3% | 17.2% |
| Walk | 0.0% | 35.0% | 7.7% | 2.9% |
| Private e-bike | 4.9% | 5.1% | 6.6% | 4.4% |
| Bike | 33.3% | 27.8% | 31.5% | 12.0% |
| Bikesharing | 1.2% | 1.2% | 2.5% | 1.2% |
| Carsharing | 0.4% | 1.0% | 1.7% | 3.9% |
| Taxi | 0.8% | 0.6% | 1.1% | 1.0% |
| Motorcycle | 3.3% | 2.8% | 3.6% | 2.4% |

The values of the *status quo* alternative attributes in the commute choice experiment are entirely provided by the respondents; moreover, if the current mode is the private car, the travel time of the shared EV alternative is constrained to be identical to the travel time of the car. If the values of the *status quo* alternative are incorrect (for example bike trips with a speed of over 40km/h or access and egress time over an hour for a short trip) it can result in unrealistic choice tasks. Therefore, we cleaned the data and only retained the responses with realistic values for the commute trip analysis. The final sample used for the commuting mode choice model has 295 respondents. Table 5 shows the attribute distribution of the final sample of current commute trips.

**Table 5 Mode attributes distribution of the reference commuting trip**

|  |  | N | Mean | Std Dev | Min | 25% | 50% | 75% | Max |
|---|---|---|---|---|---|---|---|---|---|
| Private car | Travel time (min) | 906 | 12.8 | 6.6 | 1 | 10 | 12 | 15 | 45 |
|  | Travel cost (euro) | 906 | 0.9 | 1.1 | 0 | 0.3 | 0.6 | 1.3 | 9.9 |
|  | Access and egress time (min) | 906 | 8.8 | 9.1 | 0 | 3 | 5 | 11 | 48 |
|  | Congestion frequency (%) | 906 | 28.5 | 26.8 | 0 | 10 | 20 | 50 | 100 |
|  | Congestion time (min) | 906 | 7.1 | 6.6 | 0 | 1 | 5 | 11 | 30 |
|  | Parking search time (min) | 906 | 4.2 | 4.1 | 0 | 1 | 2 | 6 | 20 |
|  | Parking fee (euro) | 906 | 4.7 | 5.5 | 0 | 0 | 3 | 8 | 20 |
| Bike | Travel time (min) | 864 | 19.2 | 8.4 | 4 | 13 | 20 | 25 | 45 |
|  | Parking search time (min) | 864 | 1.2 | 1.6 | 0 | 0 | 1 | 2 | 10 |

### 3.3 Model specification and estimation

We used mixed logit model with error components to analyse the choice data. Mixed logit model allows us to capture the unobserved correlation between the two eHUBS alternatives (shared EV and e-bike). It is estimated with a panel structure to also account for the correlation between the responses of the same individual.

Following the terminology in (De Luca and Di Pace, 2015), instead of the "holding model" which explores consumers' choices among all available transport modes, our model is a switching model which is conditional on the currently used mode: the existing mode alternative is only available for people who are currently using it. This approach does not aim to study the trade-off among all modes; instead, it focuses on whether consumers would switch from their current mode to new modes when they become available. However, our approach is slightly different from De Luca and Di Pace in that we did not estimate separate models for different current modes. Instead, we pooled all data to estimate a single model. By doing this, we can examine whether different users share some coefficients for eHUBS modes (see more details in the model specification part later).

The utility specification for the currently used mode is as follows:

$$U_{SQnt} = V_{SQnt} + \varepsilon_{SQnt} = \sum_{k=1}^{K} I_{knt}\beta_k^X X_{knt} + \varepsilon_{SQnt} \qquad (1)$$

where $U_{SQnt}$ is the utility of the current mode alternative (*Status Quo*) for respondent *n* in choice task *t*. $I_{knt}$ is an indicator which equals 1 if mode *k* is the current mode of respondent *n* in choice task *t*. $X_{knt}$ is a vector of the attributes corresponding to mode *k*. $\beta_{jk}^X$ are coefficients to be estimated. $\varepsilon_{knt}$ is an extreme value type 1 error term that is identically and independently distributed. The specification for the eHUBS modes is as below:

$$U_{jnt} = V_{jnt} + \varepsilon_{jnt} = ASC_j + \sum_{k=1}^{K} I_{knt} ASC_{jk} + \beta_{jk}^X X_{jnt} + \beta_j^S S_n + \mu_{Shared} + \mu_{jn} + \varepsilon_{jnt} \qquad (2)$$

where $ASC_{jk}$ denotes the influence of current mode *k* on the alternative specific constant of eHUBS mode *j*. $S_n$ is the vector of respondent *n*'s socio-demographic variables. Note that instead of specifying separate ASCs for each current mode in the *status quo* alternative, we estimated ASCs of the two eHUBS modes for users of each current mode ($ASC_{jk}$), which allows different relative preferences between the two shared modes to be captured. $\beta_{jk}^X$ and $\beta_j^S$ are coefficients to be estimated. There are two error components in each eHUBS alternative: $\mu_{Shared}$ captures the unobserved correlation between eHUBS alternatives and $\mu_{jn}$ is the alternative-specific error term. All error components are independently and identically distributed (with zero mean normal distribution) across respondents. The standard deviations of these error terms are also model parameters to be estimated.

We specify all attribute coefficients as alternative-specific even if the attribute is generic (such as travel cost). This alternative-specific parameter specification is widely adopted in stated/revealed preference mode choice studies and is found to increase model fit compared to generic parameter specification (Li and Kamargianni, 2020, 2018; Mo et al., 2021; Reck et al., 2020). Moreover, empirical studies have also shown that sensitivity for attributes such as cost can vary widely between different products (Erdem et al., 2002; Wang, 2012). From a theoretical point of view, this is also plausible: although the alternatives are all travel modes, the travel experience they offer are quite different (e.g. more pleasant or productive), therefore people may have different tastes for different modes which can lead to varied marginal disutility of travel time and cost (Li and Kamargianni, 2019; Shires and Jong, 2009). In the transport research this can be termed the mode-specific effect and empirical evidence has been reported in previous studies such Mackie et al. (2003) and Wardman (2004).Similar to the conditional switching model in (De Luca and Di Pace, 2015), we start with the most flexible specification and assign separate coefficients for users of each mode: for example, we estimate three separate coefficients of shared e-bike cost for people who are currently using a car, public transport and bike. If different mode users' estimates for the same attribute (e.g., the coefficient for shared e-bike cost for car and bike users) are similar in value and do not significantly reduce model fit after being constrained to be identical, they will then share a generic coefficient for a parsimonious model structure. All attribute coefficients are retained in the final model even if they do not appear to be statistically significant as suggested in (Ortúzar and Willumsen, 2011). We also tested preference heterogeneity among different socio-demographic groups. We only present the variables that have a statistically significant impact in the final model.

The probability that respondent *n* makes a sequence of choices $\boldsymbol{i} = \{i_1, \ldots, i_t\}$ is thus given by

$$P_{n\boldsymbol{i}} = \int \prod_{t=1}^{T} [\frac{e^{V_{i_t nt}}}{e^{V_{SQnt}} + \sum_j e^{V_{jnt}}}] f(\mu) d\mu$$

We estimated the model using PandasBiogeme (Bierlaire, 2020) with 1000 MLHS (modified Latin Hypercube sampling) random draws. A model with 2000 draws was also tested and the estimated coefficients are broadly similar.

# 4. Results and discussion

This section first presents the modelling results for the commuting experiment and non-commuting experiment. Then we will apply the modelling results in several scenario simulations to better understand the insights of the findings.

## 4.1 Model for non-commuting trip

The results for the choice model of the non-commuting trips are presented in Table 6. In this model we specified a distance-based ASC for walking, because the share of people switching to eHUBS becomes much higher for longer walking trips (the percentage choosing the status quo alternative drops from 80% to 55%) and this impact of distance will severely bias the travel time coefficient. Since we have mode-specific attribute coefficients, we cannot directly compare ASCs to derive mode preference. The standard deviation of the ASC of eHUBS modes are statistically significant, showing that there is preference heterogeneity for the two modes. The error component shared by the two modes is also statistically significant, indicating that people perceive the two eHUBS modes to be similar. This suggests that shared e-bike is considered to be a viable substitute for shared EV.

All travel time and cost coefficients that are statistically significant have the expected negative sign. The travel cost coefficients for car and public transport users are both statistically significant and have different values. The travel time coefficient is significant for bike and public transport users. It shall be kept in mind that existing travel modes only appear in the choice tasks for their current users and the estimates may be different from the population average. As for the non-significant travel time parameters, it is possibly due to people being indifferent within the time range or under the short trip context specified in our experiment, or that the disutility of travelling is offset by the benefits brought by conjoint activities or travel-experience factors (in which case the travel time coefficient can even become positive, see (Hess et al., 2005; Li and Kamargianni, 2018) for more discussion).

**Table 6 Results of the choice models for the non-commuting trip experiment**

| Parameters | Estimate | p-value |
|---|---|---|
| **Alternative specific constants** | | |
| Shared Electric Vehicle (SEV) | **-4.33** | 0.00 |
| SEV: shopping | **0.47** | 0.00 |
| SEV: PT user | **1.02** | 0.03 |
| SEV: Bike user | **-0.98** | 0.02 |
| SEV: Walk 2km | -0.56 | 0.44 |
| SEV: Walk 5km | 0.86 | 0.57 |
| SEV: standard deviation | **1.28** | 0.00 |
| Shared Electric Bike (SEB) | **-3.47** | 0.00 |
| SEB: shopping | -0.18 | 0.16 |
| SEB: PT user | 0.46 | 0.37 |
| SEB: Bike user | **-1.68** | 0.00 |
| SEB: Walk 2km | -0.28 | 0.71 |
| SEB: Walk 5km | 0.72 | 0.64 |
| SEB: standard deviation | **1.07** | 0.00 |
| Shared: standard deviation | **1.64** | 0.00 |
| **Mode attributes** | | |
| SQ Car: Travel time | -0.02 | 0.56 |
| SQ PT: Travel time | **-0.06** | 0.00 |
| SQ Bike: Travel time | **-0.04** | 0.01 |
| SQ Walk: Travel time | -0.01 | 0.53 |
| SQ Car: Travel cost | **-0.38** | 0.00 |
| SQ PT: Travel cost | **-0.39** | 0.01 |
| SQ All: Access/egress time | 0.02 | 0.33 |
| SQ Car: Congestion | -0.16 | 0.15 |
| SQ Car & Bike: Parking time | **-0.04** | 0.01 |
| SQ Car: Parking cost | **-0.08** | 0.02 |
| SEV: Car&PT user Travel time | **-0.06** | 0.03 |
| SEV: Bike&Walk Travel time | 0.05 | 0.21 |
| SEV: Car user Travel cost | 0.02 | 0.78 |
| SEV: PT user Travel cost | **-0.28** | 0.01 |
| SEV: Bike&Walk Travel cost | -0.16 | 0.14 |
| SEV: All user Access/egress time | **-0.04** | 0.00 |
| SEV: Car user Congestion | **-0.37** | 0.00 |
| SEV: PT,Bike&Walk user Congestion | 0.02 | 0.81 |
| SEB: Car&PT user Travel time | **-0.05** | 0.00 |
| SEB: Bike&Walk Travel time | 0.02 | 0.38 |
| SEB: All user Travel cost | **-0.47** | 0.00 |
| SEB: All user Access/egress time | **-0.03** | 0.02 |
| **Socio-demographic variables** | | |
| SEV: age<=35 | **1.45** | 0.00 |
| SEV: with children | **1.48** | 0.00 |
| SEB: age<=35 | **0.68** | 0.00 |
| SEB: age>=60 | **-0.99** | 0.00 |
| SEB: income<=40.000 euro | **0.48** | 0.01 |
| SEB: higher education | **0.46** | 0.02 |
| SEB: with children | **0.72** | 0.00 |
| **Model summary** | | |
| Null LL | -4660.3 | |
| Final LL | -2354.4 | |
| pseudo rho square | 0.495 | |
| Number of model parameters | 44 | |
| Number of observations | 4242 | |
| Number of individuals | 867 | |

We will focus on the eHUBS cost and time coefficients. Users of all modes share a generic cost coefficient for shared e-bike, while this is not the case for shared EV: cost only has a significant impact for public transport users. As for the travel time coefficients, users of the two slow modes (bike and walk) are not significantly affected by the travel time of the two eHUBS modes, probably because these two modes are much faster than their current mode and the variation in time does not make a big difference. The travel time of the shared modes is significant for users of cars and public transport. For public transport users (because only they have a significant cost coefficient for both shared modes), the corresponding VOT is respectively 6.4 and 12.9 euro/hour for shared e-bike and EV.

We estimated a generic coefficient for access time for all mode users and it was found to significantly affect the intended usage of both eHUBS modes. However, its absolute value is lower than that of the travel time coefficient for car and public transport users. This is probably because car and public transport users' travel time sensitivity is higher than that of people who walk and cycle. In a model with generic eHUB mode travel time coefficients for the entire population, we indeed found the absolute value of the access time coefficient to be larger than that for travel time. The access time for the current mode is not significantly different from zero.

Moreover, higher parking costs and longer parking search time are both found to significantly reduce the use of the current mode (private car and bike), suggesting that higher parking pressure and costs of the private car can encourage car users to switch towards shared modes.

The congestion variable in the non-commuting model is specified as the probability of congestion multiplied by the ratio of possible congestion time to travel time.[2] Car users are less likely to use shared EV when there is more congestion during the trip, while users of other modes are not found to be significantly deterred by a higher extent of congestion. This is probably because people, in general, care less about congestion in the context of non-commuting trips which are usually not with high time pressure (Wardman and Nicolás Ibáñez, 2012), but car users prefer to stick to the comfort of their own private car when there is more congestion (the congestion extent is the same between shared EV and private car in the experiment).

---

[2] We also tested the congestion specification used in the commuting model. The model fit (log-likelihood) is slightly lower, therefore we adopted the specification described in the text.

As for trip purposes, we found that people are more likely to choose shared EV for shopping trips (in contrast to leisure trips). In the case of the shared e-bike, the coefficient for a shopping trip is negative but only statistically significant at a 90% confidence level, which suggests that given their capabilities of carrying large goods, e-cargo bikes still cannot substitute as many shopping trips as e-bikes in the case of leisure trips. A larger sample is needed to be conclusive regarding this finding.

As for the impact of socio-demographic variables, the impact of gender is non-significant in the choice of using eHUBS for non-commuting trips. Young people are more willing to use the eHUBS modes, while older people are less inclined to use e-bike probably due to worse health conditions. Bielinski et al. (2021) mentioned in their literature review that several previous studies found that privately-owned e-bike has considerable popularity among older people (above 50-55 years old), but their study also found that older people are less likely to use shared e-bikes. Low-income earners are more likely to choose shared e-bike, indicating its potential in increasing the accessibility of this group. The impact of high education is positive and the coefficient for shared e-bike is statistically significant. People with children are significantly more inclined to use both shared modes. This resonates with the fact that many Dutch carsharing users are people who have families with small children (KiM, 2015) and also has been found in studies conducted in Norway (Mouratidis, 2022).

### 4.2 Model for commuting trip

Table 7 presents the results of the mixed logit model for commuting trip. Since the sample used for the estimation of the commuting trip model is much smaller than the model for non-commuting trips, these results are exploratory and mostly used to detect whether there are major differences compared to non-commuting trips. Similar to the non-commuting model, the common error component for shared mobility in the mixed logit model is statistically significant, showing that the perceived similarity between eHUBS modes also holds in the case of commuting trips.

All travel time and cost coefficients have a negative sign except the travel time coefficient of the private car (although non-significant). While interpreting the results, it shall be kept in mind that the attributes of the current mode alternative are not varied within-person unlike the two eHUBS modes. For each individual, the attribute values of the current mode are fixed throughout the experiment, which strictly speaking does not allow within-person comparison such as "to what extent can variation of travel time change the probability of choosing the alternative", although it is in general acceptable if these variables are considered exogenous (Chorus and Kroesen, 2014). Moreover, the existing modes only appear in the choice tasks for their current users which may render the attribute coefficient estimates different from the population average. Apart from travel time and cost attributes, we also examined attributes related to parking of private cars and bikes: parking cost of private cars is non-significant, while parking search time only has a significant impact on bike users, indicating that bike users with more parking difficulty are more likely to switch towards eHUBS modes.

We will again focus on the travel time and cost coefficients of the two eHUBS modes (shared e-bike and shared EV). Users of both current modes share a generic cost coefficient for shared e-bike, while the coefficients for shared EV cost are significantly different between users of private cars and bikes, although both turn out to be non-significant in the final model.

**Table 7 Results of the choice models for the commuting trip experiment**

| Parameters | Estimate | p-value | Estimate | p-value |
|---|---|---|---|---|
| **Alternative specific constants** | | | | |
| Shared Electric Vehicle (SEV) | **-4.46** | 0.00 | **-4.26** | 0.00 |
| SEV: Bike user | -0.61 | 0.57 | -0.74 | 0.52 |
| SEV: standard deviation | **-1.11** | 0.00 | **0.95** | 0.00 |
| Shared Electric Bike (SEB) | **-4.15** | 0.00 | **-4.20** | 0.00 |
| SEB: Bike user | -0.79 | 0.42 | -0.84 | 0.44 |
| SEB: standard deviation | **0.85** | 0.00 | 0.08 | 0.91 |
| Shared: standard deviation | **2.28** | 0.00 | **2.23** | 0.00 |
| **Mode attributes** | | | | |
| Status Quo (SQ) Car: Travel time | 0.01 | 0.79 | 0.03 | 0.38 |
| SQ Bike: Travel time | -0.03 | 0.49 | -0.02 | 0.52 |
| SQ All: Travel cost | **-0.44** | 0.01 | **-0.69** | 0.03 |
| SQ All: Access/egress time | 0.00 | 0.89 | 0.00 | 0.95 |
| SQ Car: Congestion | 0.02 | 0.87 | 0.03 | 0.79 |
| SQ Car: Parking search time | *-0.16* | 0.06 | -0.16 | 0.11 |
| SQ Bike: Parking search time | **-0.79** | 0.00 | **-0.72** | 0.00 |
| SQ Car: Parking cost | -0.08 | 0.12 | -0.07 | 0.16 |
| SEV: Car user Travel time | *-0.08* | 0.09 | *-0.08* | 0.10 |
| SEV: Bike user Travel time | **-0.20** | 0.02 | *-0.19* | 0.09 |
| SEV: Car user Travel cost | 0.11 | 0.26 | 0.11 | 0.20 |
| SEV: Bike user Travel cost | **-0.42** | 0.01 | -0.36 | 0.32 |
| SEV: All user Access/egress time | -0.03 | 0.11 | -0.02 | 0.21 |
| SEV: Car user Congestion | 0.05 | 0.42 | 0.05 | 0.38 |
| SEV: Bike user Congestion | *-0.22* | 0.07 | -0.08 | 0.74 |
| SEB: All user Travel time | **-0.04** | 0.04 | -0.04 | 0.12 |
| SEB: All user Travel cost | **-0.46** | 0.00 | **-0.35** | 0.03 |
| SEB: Car user Access/egress time | 0.03 | 0.25 | 0.03 | 0.27 |
| SEB: Bike user Access/egress time | **-0.07** | 0.00 | *-0.04* | 0.09 |
| **Socio-demographic variables** | | | | |
| SEV: age<=35 | **1.96** | 0.00 | **1.79** | 0.00 |
| SEV: income>=80.000 euro | **-1.75** | 0.00 | **-1.85** | 0.03 |
| SEB: age<=35 | **0.99** | 0.02 | **1.10** | 0.01 |
| SEB: age>=60 | **-3.32** | 0.01 | **-1.91** | 0.04 |
| SEB: income>=80.000 euro | **-1.49** | 0.01 | **-1.59** | 0.04 |
| **Model summary** | | | | |
| Null LL | -1944.5 | | -1944.5 | |
| Final LL | -708.7 | | -708.7 | |
| pseudo rho square | 0.636 | | 0.636 | |
| Number of model parameters | 31 | | 31 | |
| Number of observations | 1770 | | 1770 | |
| Number of individuals | 295 | | 295 | |

The travel time coefficient for the shared e-bike mode is also generic: it was statistically significant in the multinomial logit model but turned non-significant in the final mixed logit model, probably due to the small sample size. It corresponds to the value of time (VOT) of around 11.4 euro/hr which is higher than the value in non-commuting trip as expected (Shires and Jong, 2009). As for the travel time coefficient for shared EV, we kept the coefficients separate for different mode users. The travel time coefficient for bike users is extremely large (corresponding to a VOT of over 31.7 euro/hour) and statistically significant. Instead of revealing the genuine impact of travel time, this most likely indicates the impact of distance in our sample: there are only around 30 respondents choosing shared EV among bike users and they are almost all distributed among trips shorter than 6km, therefore trips with longer distance (and longer travel time) are greatly penalized. Since the number of responses choosing shared EV among bike users is too few, we did not add a distance-specific ASC to correct this bias. As for the access time of shared e-bike, bike users value it around the same as travel time, while we do not find its impact to be significant for car users. Access time for shared EV is non-significant as well. Congestion is specified as possible congestion time multiplied by the probability of congestion and it is found to be non-significant for both car and bike users.

Similar to what happens with non-commuting trips, young age is also a significant indicator for the usage of eHUBS for non-commuting trips, while the impact of gender is non-significant. Older people are still less inclined to use shared e-bike. People with high income are significantly less likely to switch to eHUBS modes, while low income also has a significant negative impact on the use of shared EV. Although we tested the interaction of income and cost and found it to be non-significant, maybe carsharing is in general too expensive for people with a low income. Unlike the case of non-commuting trips, highly educated people are not found to be more likely to switch to eHUBS for their commute trips. In general, the results regarding socio-demographic variables for both commuting and non-commuting trips match the typical early adopter profile and confirm most related findings on transport innovation adoption (Becker et al., 2017; Campbell et al., 2016; Hess and Schubert, 2019; Hu et al., 2018; Jie et al., 2021; Kaplan et al., 2018; Wang and Yan, 2016).

## 4.3 Mode substitution and policies promoting eHUBS use for car users

In order to extract more tangible insights from the model, we conducted several numerical simulations to illustrate our findings. We would like to stress again that the main purpose of this article is not forecasting market share under constraints in reality and the choice probabilities in the simulation are used for comparison between different scenarios. Many studies based on stated preference data conducted similar simulations analysis to provide insights regarding the impact of attribute variation or policy incentives (Alonso-González et al., 2020; Danielis et al., 2020; Hackbarth and Madlener, 2016, 2013; Tanaka et al., 2014).

First, we will simulate the choice probability of each alternative for users of each mode under different distances. This allows us to explore the eHUBS preference of different modes and also observe how they vary by trip distance. Second, to identify effective measures for promoting private car users to switch towards eHUBS, we will also simulate the impact of several policies on car users' choice probability.

### 4.3.1 Mode substitution patterns of users of different modes

We used the non-commuting choice model estimates because the reference trip is also varied in the non-commuting experiment which allows for within-person comparison. We simulate for leisure trips and the attribute values are all set to be the middle level in the experiment setting (apart from zero congestion): this level represents the average condition of these modes and can serve as a reference point. The individual is assumed to be a young person with high education, which is supposedly the most inclined to use eHUBS.

Table 8 presents the mode substitution pattern of users of each mode under three different distance contexts. We can see that the mode substitution pattern differs greatly by people's current mode and trip distance. Regarding the influence of distance, we observe that the share of switchers among active mode users spike for long-distance trips (walkers at 5km and bike users at 10km). Public transport users have a slightly stronger preference for eHUBS modes for longer distances, which is noticeably mainly caused by the increased share of shared e-bike due to its larger relative advantage in travel time and cost for long distance trips. The share of switchers of car users is relatively stable across all distances.

As for the relative preference between the two shared modes in the eHUBS, public transport users and walkers have a stronger preference for shared e-bikes, while car and cyclists have a higher preference for shared EVs. The fact that cyclists prefer EV to e-bike is at odds with the conclusion of (Mo et al., 2021) that people are more likely to switch to a new mode if it is more similar to their current mode. In our case it seems to be the other way around: if the new mode is too similar to the currently used mode and does not provide sufficient added value, there is less need to switch. This is probably why the share of cyclists choosing shared e-bike increases with trip distance: the benefits of e-bike (higher speed and less effort) are more pronounced on long trips.

In general, car users are less likely to switch towards eHUBS modes compared to public transport users. Active mode users' inertia for their current mode is on par with car users for short trips, but their preference for eHUBS strongly increases with distance.

**Table 8 Simulation results of mode substitution for non-commuting trips of young people with higher education**

| Current mode | 2km Status Quo | Shared EV | Shared e-bike | 5km Current mode | Shared EV | Shared e-bike | 10km Current mode | Shared EV | Shared e-bike |
|---|---|---|---|---|---|---|---|---|---|
| Car | 75.6% | 12.2% | 12.3% | 77.7% | 11.0% | 11.2% | 78.7% | 11.2% | 10.1% |
| PT | 68.1% | 16.8% | 15.1% | 67.5% | 14.6% | 17.9% | 59.2% | 13.6% | 27.2% |
| Bike | 86.1% | 8.6% | 5.3% | 82.4% | 10.3% | 7.4% | 70.8% | 15.5% | 13.6% |
| Walk | 78.7% | 8.7% | 12.5% | 56.0% | 20.2% | 23.8% | | | |

Table 9 shows the simulation results for commuting trips. It uses identical attribute values as the non-commuting trip simulation (apart from a zero parking search time for bike users because the vast majority of bike users in our sample has 0 or 1 minute of parking search time for their commuting trip). We exclude the results of 10km trips for bike users because we hardly have any respondents with such commuting trips in our sample. Several observations can be made: Firstly, young car users are much more willing to switch towards eHUBS for commuting trips compared to non-commuting trips. Secondly, bike users in general stick to cycling for commuting trips.

**Table 9 Simulation results of mode substitution for commuting trips of young people**

| Current mode | 2km Status Quo | Shared EV | Shared e-bike | 5km Status Quo | Shared EV | Shared e-bike | 10km Status Quo | Shared EV | Shared e-bike |
|---|---|---|---|---|---|---|---|---|---|
| Car | 66.4% | 22.6% | 11.1% | 66.2% | 20.9% | 12.8% | 67.6 % | 19.8% | 12.6% |
| Bike | 90.5% | 5.4% | 4.2% | 93.4% | 2.1% | 4.5% | | | |

### 4.3.2 The impact of policies on mode substitution patterns of car users

We tested the impact of four policies/measures on reducing car use and promoting eHUBS use. Two of them focus on increasing the attractiveness of eHUBS by reducing its access time and reducing the usage fee. We did not test reduction on shared EV price because car users are not sensitive to its cost variation. The other two policies aim to curb car use by increasing parking search time (for example by reducing available parking space) and raising parking fees.

The simulation uses a 5km non-commuting trip as the base scenario. Table 10 presents the results. The effectiveness of cutting eHUBS access time is almost on par with the two policies targeting car use, which probably makes it a less efficient measure: reducing access time implies expanding the network and increasing the density of eHUBS, which requires investment in new vehicles and hub facilities and can be rather costly. Moreover, this measure would also induce more public transport and active modes users to switch towards shared EV which is not a desirable effect. Reducing the usage fee of shared e-bike seems to be the most effective in terms of reducing car use and increasing shared e-bike usage. Moreover, it even reduced use of shared EV compared to base scenario due to the substitution effect between shared e-bike and EV. However, it is at the cost of harming the profitability and sustainability of shared mobility providers. The findings of this simulation exercise seem to suggest that eHUBS deployment can be accompanied by policies targeting car use to realize its potential in cutting car use and its externalities.

**Table 10 Simulation results of the impact of different policies on mode substitution of car users**

| Policy | Status Quo | Shared EV | Shared e-bike |
|---|---|---|---|
| Base | 77.7% | 11.0% | 11.2% |
| Access time to eHUBS from 10 to 7.5 minutes | 76.6% | 11.7% | 11.7% |
| SEB price from €1 to €0.5 | 76.2% | 10.6% | 13.2% |
| Car parking search time from 5 to 7.5 minutes | 76.3% | 11.7% | 12% |
| Car parking fee from €3 to €4.5 | 76.4% | 11.7% | 11.9% |

## 5. Conclusion

This paper studies people's mode choice when eHUBS become available. More specifically, we investigate people's choice regarding whether to use shared electric mobility services including shared EV and e-bike to substitute their current modes (private car, public transport and active modes). We also explicitly study how the preference for and choice of using eHUBS varies for users of those different current modes. We used a stated choice experiment which is tailor-made for each current mode user: the choice set consists of only the two eHUBS modes and the respondent's current mode for a trip of a certain distance and purpose. We estimated mixed logit models to study people's choice of using eHUBS modes to substitute their current modes. We also conducted several simulations based on the results of the choice models to give insights regarding the mode substitution patterns and effectiveness of various policies in reducing private car use and promoting shared mobility.

We explored the unobserved correlation between shared EV and e-bike by assigning a communal error component for the two eHUBS modes. The results show that the standard deviation of the component is statistically significant in both models for commuting and non-commuting trips, which suggests that in both contexts these two modes are perceived to be similar.

We found that most factors which are influential in choices among existing modes are also found to play a significant role in the choice of eHUBS adoption, such as transport mode attributes (travel time and cost, access time, parking search time and parking fee for private car) and socio-demographic variables. Apart from different general preferences for shared electric modes, our choice models also show that users of different current modes have different sensitivity for eHUBS attributes such as travel time and travel cost. This further confirms the necessity of investigating the impact of current transport modes on people's choice regarding new transport services.

The simulation based on the choice model results show that under average trip conditions (in terms of attribute values), a small but significant share of young people with higher education are willing to use eHUBS to replace their current travel mode. This result holds for both commuting and non-commuting trip, which demonstrates the potential of eHUBS in replacing car trips for both trip purposes. As for non-commuting trips, shared EVs are preferred for shopping trips compared to leisure trips, while the preference for using e-bikes in leisure trips is higher than the preference for using e-cargo bikes for shopping trips, although the latter effect is only statistically significant at 90% confidence level and calls for more evidence to be conclusive.

We also found that the substitution patterns differ greatly by current mode and trip distances. Given similar trip conditions (in terms of the attributes of current modes), public transport users are consistently more willingly to switch towards eHUBS compared to car users. Active modes users have even stronger inertia for their current modes (bike and walking) than car users for short trips (shorter than 2km for walking and 5km for bike), while their preference for shared modes increases when the trip is longer than the threshold. Our findings can enhance our understanding of people's adoption of eHUBS modes, which can play a vital role in the design, deployment and promotion of eHUBS services.

Our study has two main limitations. A larger sample from the commuting experiment would have allowed us to achieve more reliable estimates for the corresponding mode choice model. We also only allowed respondents to specify a single (most frequently used) mode for each trip while people can be multimodal; this may have an impact on our results considering that preferences vary between different mode users.

There remain many areas for future research regarding the mode choice involving shared electric modes and eHUBS. First, since public transport and active modes already take a significant share in Amsterdam, similar studies can be conducted in other cities which are more car-dominated to see whether car users in these places have different preferences for shared modes. Second, our study focuses on the usage of one-way station-based eHUBS for short full trips and assumes full availability of shared vehicles. It is interesting to investigate how different types of organizations (e.g. roundtrip, free-floating), different trip contexts (first/last-mile connecting trips combined with public transport), and uncertain and limited availability affect people's preference for shared mobility services. Furthermore, it is also crucial to study the added-value of mobility hubs: is integrating multiple shared modes more effective in boosting shared mode usage and reducing car use compared to independent existing single-mode shared mobility services? Finally, when more eHUBS and shared electric mobility services are available, it is important to use revealed preference data to assess the real impact of eHUBS and the validity of stated preference studies.

**Author contributions**

FL: Conceptualization, Experiment design, Data collection, Statistical analysis, Writing—original draft. GB, DD, NT, MB, GC: Experiment design. All authors reviewed the manuscript.

**Acknowledgments**

This work was supported by the eHUBS - Smart Shared Green Mobility Hubs project funded by Interreg.

**Declarations of interest**

None.